\documentclass[12pt]{article}
\usepackage{a4wide,epsfig,psfrag,amsmath,amssymb,cite,scalefnt}
\usepackage{color}
\usepackage{amsmath,comment,braket}
\usepackage{placeins}
\usepackage{subcaption}

\graphicspath{ {figs_ggzh0/} }

\newcommand{\be}{\begin{equation}}
\newcommand{\ee}{\end{equation}}
\newcommand{\bea}{\begin{eqnarray}}
\newcommand{\eea}{\end{eqnarray}}

\allowdisplaybreaks

\newcommand{\gsim}{\;\rlap{\lower 3.5 pt \hbox{$\mathchar \sim$}} \raise 1pt
 \hbox {$>$}\;}
\newcommand{\lsim}{\;\rlap{\lower 3.5 pt \hbox{$\mathchar \sim$}} \raise 1pt
 \hbox {$<$}\;}

\begin{document}

\title{\vskip-3cm{\baselineskip14pt
    \begin{flushleft}
      \normalsize TTP20-041, P3H-20-074
    \end{flushleft}}
  \vskip1.5cm
  Virtual corrections to $gg\to ZH$ in the high-energy and large-$m_t$ limits
}

\author{
  Joshua Davies$^{a}$,
  Go Mishima$^{b}$,
  Matthias Steinhauser$^{b}$
  \\[1mm]
  {\small\it $^a$ Department of Physics and Astronomy, University of Sussex,
    Brighton BN1 9QH, UK}
  \\[1mm]
  {\small\it $^b$ Department of Physics, Tohoku University, Sendai, 980-8578 Japan}
  \\[1mm]
  {\small\it $^c$ Institut f{\"u}r Theoretische Teilchenphysik}\\
  {\small\it Karlsruhe Institute of Technology (KIT)}\\
  {\small\it Wolfgang-Gaede Stra\ss{}e 1, 76128 Karlsruhe, Germany}
}
  
\date{}

\maketitle

\thispagestyle{empty}

\begin{abstract}

  We compute the next-to-leading order virtual corrections to the partonic
  cross-section of the process $gg\to ZH$, in the high-energy and large-$m_t$
  limits.  We use Pad\'e approximants to increase the radius of convergence of
  the high-energy expansion in $m_t^2/s$, $m_t^2/t$ and $m_t^2/u$ and show
  that precise results can be obtained down to energies which are fairly close
  to the top quark pair threshold. We present results both for the form
  factors and the next-to-leading order virtual cross-section.

\end{abstract}

\thispagestyle{empty}

\sloppy


\newpage


\section{Introduction}

At the CERN Large Hadron Collider (LHC), gluon fusion processes play an
important role due to the large gluon luminosities at high collision
energies. As a consequence one often observes that gluon fusion--induced
processes provide a numerically large contribution to the theory predictions of
production cross-sections.  This is true even for processes for which the
leading-order (LO) contribution consists of one-loop diagrams.  A prime
example of such a process is inclusive Higgs boson production, where the
gluon-fusion channel is about an order of magnitude larger than all other
contributions.

In this paper we consider the associated production of a $Z$ and a Higgs
boson, $pp\to ZH$, often called ``Higgs Strahlung''.  This process
was of particular importance in the observation of the Higgs boson's decay into
bottom quarks at ATLAS~\cite{Aaboud:2018zhk} and CMS~\cite{Sirunyan:2018kst}.
At LO it is mediated by a tree-level process in which a quark and anti-quark
annihilate. For this channel, corrections up to next-to-next-to-next-to-leading
(N$^3$LO) order are available~\cite{Kumar:2014uwa} (see also
Ref.~\cite{Heinrich:2020ybq} and references therein) and are included in the
program {\tt vh@nnlo}~\cite{Brein:2012ne,Harlander:2018yio}.  Electroweak and
QCD corrections are also included in the program {\tt HAWK}~\cite{Denner:2014cla}.

Associated $ZH$ production can also occur via the loop-induced 
gluon fusion process. Although formally of next-to-next-to-leading order
(NNLO) with respect to $pp\to ZH$, it provides a sizeable contribution,
in particular in the 
boosted Higgs boson regime in which the transverse momentum of the Higgs boson
is large~\cite{Englert:2013vua,Harlander:2013mla}. Furthermore,
the process $gg\to ZH$ provides sizeable contributions to the uncertainties
of $ZH$ production with a subsequent decay of the Higgs boson in a pair of
bottom quarks, e.g., Ref.~\cite{Aad:2020jym}.
Being loop-induced, $gg\to ZH$ is sensitive to physics beyond the
Standard Model. In Ref~\cite{Harlander:2018yns} it was suggested that
the gluon-initiated component of $pp\to ZH$ can be extracted by comparing to
$WH$ production, which only has a Drell-Yan--like component.
It is thus important to consider NLO QCD corrections to $gg\to ZH$,
requiring the computation of two-loop box-type Feynman diagrams
with two different final-state masses ($m_Z$ and $m_H$) and the massive top
quark propagating in the loops.

An exact LO (one-loop) calculation was performed in \cite{Kniehl:1990iva}.
At NLO only approximations in the large $m_t$ limit are
known~\cite{Altenkamp:2012sx,Hasselhuhn:2016rqt}.  In this work we consider
the two-loop NLO virtual corrections in the high-energy limit, expanding the
two-loop master integrals for $m_t^2 \ll s,t$, where $s$ and $t$ are the
Mandelstam variables. Furthermore, we also provide analytic results for the form
factors in the large-$m_t$ limit.\footnote{In~\cite{Hasselhuhn:2016rqt}
  only the squared amplitude has been computed.}
A similar approach has been applied to the related process of Higgs boson pair
production, $gg\to HH$, where a comparison to exact numerical calculations~\cite{Borowka:2016ypz}
could be performed and good agreement was found, even for relatively small
values for the Higgs boson transverse momentum~\cite{Davies:2019dfy}.
In Ref.~\cite{Davies:2020lpf} the high-energy expansion was successfully
applied to the top quark contribution of the two-loop diagrams
contributing to $gg\to ZZ$.

The remainder of the paper is organized as follows: In
Section~\ref{sec::notation} we introduce our notation, and give our
definitions for the form factors and the virtual finite cross-section.  In
Section~\ref{sec::1l-comparison} we consider the quality of our approximations
by comparing with the exact LO expressions.  In Section~\ref{sec::lme}, we
briefly discuss the two-loop form factors in the large-$m_t$ limit and in
Section~\ref{sec::highenergy} we discuss the form factors in the high-energy
limit, and investigate the behaviour of Pad\'e approximants constructed using
this expansion.  In Section~\ref{sec::vfin} we study the NLO virtual finite
cross-section and apply our Pad\'e scheme to extend the approxmation to a
larger kinematic region.  Finally in Section~\ref{sec::concl} we conclude our
findings. Auxiliary material can be found in the Appendix.  In
Appendix~\ref{app::1PR} we present analytic results for the one-particle
reducible double-triangle contribution and in Appendix~\ref{app::projg5} we
briefly discuss out treatment of $\gamma_5$ and the application of projectors
to obtain the form factors. In Appendix~\ref{app::helamp} we present the
relations between the form factors and helicity amplitudes for the process
$gg\to ZH$.  The expansions of the form factors are provided in an analytic,
computer readable form in the ancillary files of this paper~\cite{progdata}.


\section{Notation and technicalities}
\label{sec::notation}

We consider the amplitude for the process $g(p_1)g(p_2)\to Z(p_3)H(p_4)$ where
all momenta are assumed to be incoming.  This leads to the following definitions for the
Mandelstam variables,
\begin{eqnarray}
\label{eq::mandelstam}
  s &=& (p_1+p_2)^2\,,\nonumber\\
  t &=& (p_1+p_3)^2\,,\nonumber\\
  u &=& (p_1+p_4)^2\,,
\end{eqnarray}
with
\begin{eqnarray}
  s + t+ u &=& m_Z^2 + m_H^2\,.
\end{eqnarray}
Additionally, $p_1^2=p_2^2=0$, $p_3^2=m_Z^2$ and $(p_1+p_2+p_3)^2=m_H^2$.
We also introduce the transverse momentum ($p_T$) and
the scattering angle ($\theta$) of the final-state bosons, which are related to the
Mandelstam variables as follows,
\begin{align}
  t &=-\frac{s}{2}\left(1 - \beta \cos \theta \right) + \frac{m_H^2+m_Z^2}{2}\,, \nonumber\\
  u &=-\frac{s}{2}\left(1 + \beta \cos \theta \right) + \frac{m_H^2+m_Z^2}{2}\,, \nonumber\\
  p_T^2 &=\frac{u t-m_H^2 m_Z^2}{s}= \frac{s}{4} \beta^2 \sin ^2 \theta  \,,
\end{align}
where
\begin{align}
\beta&=
\sqrt{
1-2\frac{m_{Z}^{2}+m_{H}^{2}}{s}
+\frac{\left(m_{Z}^{2}-m_{H}^{2}\right)^{2}}{s^2}
}
\,.
\end{align}

We denote the polarization vectors of the gluons and the $Z$ boson by
$\varepsilon_{\lambda_1,\mu}(p_1)$, 
$\varepsilon_{\lambda_2,\nu}(p_2)$ and
$\varepsilon_{\lambda_3,\rho}(p_3)$,
in terms of which the amplitude can be written as
\begin{eqnarray}
\label{eq::helicity_amp}
  \mathcal{M}_{\lambda_1,\lambda_2,\lambda_3} &=& A^{\mu\nu\rho}\:
               \varepsilon_{\lambda_1,\mu}(p_1)\:
               \varepsilon_{\lambda_2,\nu}(p_2)\:
               \varepsilon_{\lambda_3,\rho}(p_3)\,.
\end{eqnarray}

Due to charge-conjugation invariance, the vector coupling of the $Z$
boson to the quarks in the loop does not contribute. The axial-vector coupling
is proportional to the third component of the isospin and thus
mass-degenerate quark doublets also do not contribute. Since we consider the lightest
five quarks to be massless, only contributions from the top-bottom doublet remain.
Furthermore, 
each individual term of $A^{\mu\nu\rho}$ is proportional to the totally anti-symmetric
$\varepsilon$ tensor from the axial-vector coupling.

\begin{figure}[t]
  \centering
  \includegraphics[width=\textwidth]{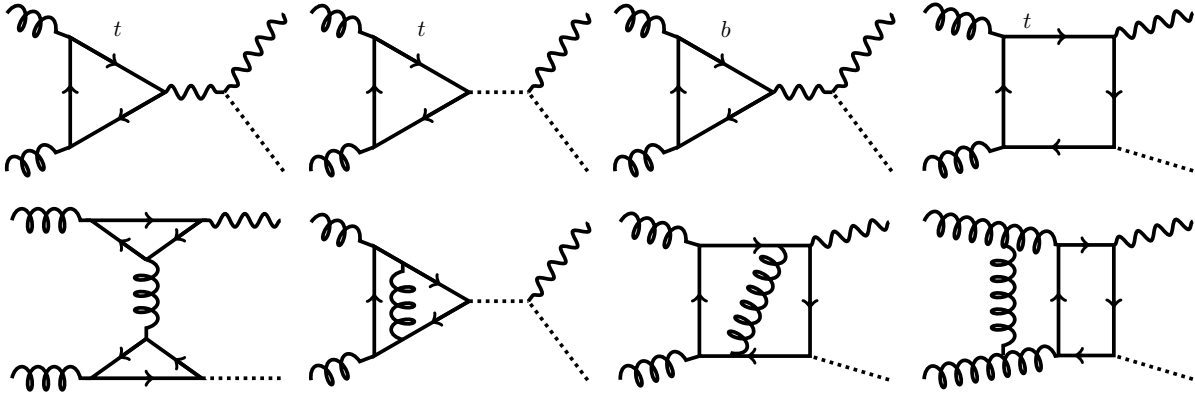}
  \caption{\label{fig::diags}LO and NLO Feynman diagrams contributiong to
    $gg\to ZH$. Curly, wavy and dashed lines represent gluons, $Z$ bosons and
    scalar particles (Higgs or Goldstone bosons), respectively. Solid lines
    stand for top or bottom quarks. Note that the latter are only present in
    the triangle diagrams.}
\end{figure}

At LO and NLO both triangle- and box-type diagrams have to be
considered. Examples of these diagram classes are shown in
Fig.~\ref{fig::diags}.  In the box-type diagrams the Higgs boson couples
directly to the quark loop, so diagrams involving the bottom quark are
suppressed by their Yukawa coupling with respect to diagrams involving the top
quark.  This is not the case for the triangle-type diagrams; here
contributions from both the top and bottom quark loops must be considered.  At
NLO there is also the contribution from the one-particle reducible
double-triangle contribution, shown as first diagram in the second row of
Fig.~\ref{fig::diags}. Note that in the numerical results discussed in the
main part of this paper these contributions are excluded, however, we present
exact analytical results in Appendix~\ref{app::1PR}. The contribution from the
reducible double-triangle diagrams to the (differential) partonic cross-
section is implemented in the computer program which comes together with
Ref.~\cite{Hasselhuhn:2016rqt}.

In total one can form 60 tensor structures from
the indices $\mu,\nu,\rho$, the independent momenta $p_1, p_2$ and $p_3$ and 
an $\varepsilon$ tensor.
Details of the computation of these 60 structures and our treatment of
$\gamma_5$ in $d=4-2\epsilon$ dimensions are given in Appendix~\ref{app::projg5}.
After applying
transversality conditions ($p_i\cdot \varepsilon_{\lambda_i}(p_i) =0$),
gauge invariance w.r.t.\ the gluons ($p_{1\mu}A^{\mu\nu\rho} = p_{2\nu}A^{\mu\nu\rho} = 0$)
and Bose symmetry ($A^{\mu\nu\rho}(p_1,p_2,p_3) = A^{\nu\mu\rho}(p_2,p_1,p_3)$),
14 tensor structures remain which can be grouped such that one has
to introduce 7 form factors.
We follow the decomposition of Ref.~\cite{Kniehl:1990iva} and write
\begin{eqnarray}
  \lefteqn{
  A_{ab}^{\mu\nu\rho}(p_1,p_2,p_3) 
  = i \delta_{ab} \frac{ \sqrt{2}G_F M_Z }{s} \frac{\alpha_s(\mu)}{\pi}
  \tilde{A}^{\mu\nu\rho}(p_1,p_2,p_3)\,, }
      \nonumber\\
  \lefteqn{\tilde{A}^{\mu\nu\rho}(p_1,p_2,p_3) =}\nonumber\\
  &&    \Bigg\{ \left( \frac{s}{2}\varepsilon^{\mu\nu\rho\alpha} p_{2\alpha} 
      - p_2^\mu \varepsilon^{\nu\rho\alpha\beta} p_{1\alpha} p_{2\beta}
      \right) F_1(t,u)
      - 
      \left( \frac{s}{2}\varepsilon^{\mu\nu\rho\alpha} p_{1\alpha} 
      - p_1^\nu \varepsilon^{\mu\rho\alpha\beta} p_{1\alpha} p_{2\beta}
      \right) F_1(u,t)
      \nonumber\\&&\mbox{}
      + \left(p_3^\mu + \frac{m_Z^2-t}{s} p_2^\mu\right)
        \varepsilon^{\nu\rho\alpha\beta} p_{2\alpha}
        \Big[ p_{1\beta} F_2(t,u) + p_{3\beta} F_3(t,u) \Big]
      \nonumber\\&&\mbox{}
      + \left( p_3^\nu + \frac{m_Z^2-u}{s} p_1^\nu\right)
        \varepsilon^{\mu\rho\alpha\beta} p_{1\alpha}
        \Big[ p_{2\beta} F_2(u,t) + p_{3\beta} F_3(u,t) \Big]
      \nonumber\\&&\mbox{}
      + \left(\frac{s}{2} \varepsilon^{\mu\nu\rho\alpha} p_{3\alpha}
             -p_2^\mu  \varepsilon^{\nu\rho\alpha\beta} p_{1\alpha}p_{3\beta}
             +p_1^\nu  \varepsilon^{\mu\rho\alpha\beta} p_{2\alpha}p_{3\beta}
             +g^{\mu\nu} \varepsilon^{\rho\alpha\beta\gamma} p_{1\alpha}p_{2\beta}p_{3\gamma}
        \right) F_4(t,u)
                            \Bigg\},
\label{eq::Amunuro}
\end{eqnarray}
where $a,b$ are colour indices.
Note that while the decomposition is the same, the form factors in
Ref.~\cite{Kniehl:1990iva} have dimension $1/\mbox{GeV}^2$ whereas we pull out an
overall factor of $1/s$ such that our form factors are dimensionless.
Only $F_1$ receives contributions from the triangle-type diagrams discussed above.
We represent the expansion coefficients of the form factors in the strong coupling
constants with the following notation,
\begin{eqnarray}
	F &=& F^{(0)} + \frac{\alpha_s(\mu)}{\pi} F^{(1)} + \cdots \,.
	\label{eq::FasExp}
\end{eqnarray}

At this point it is convenient to 
define new form factors which are linear combinations of those of
Eq.~(\ref{eq::Amunuro}),
\begin{eqnarray}
  F_{12}(t,u) &=& F_1(t,u) - \frac{t-m_Z^2}{s} F_2(t,u)\,,\nonumber\\
  F_{12}(u,t) &=& F_1(u,t) - \frac{u-m_Z^2}{s} F_2(u,t)\,,\nonumber\\
  F_2^-(t,u) &=& F_2(t,u) - F_2(u,t)\,, \nonumber\\
  F_2^+(t,u) &=& F_2(t,u) + F_2(u,t)\,.
  \label{eq::F12_F-+}
\end{eqnarray}
It is easy to see that $F_2^+(t,u)$ drops out in the squared
amplitude and thus does not contribute to physical quantities.
It is furthermore convenient to introduce
    \begin{eqnarray}
      F_{12}^-(t,u) &=& F_{12}(t,u) - F_{12}(u,t)\,, \nonumber\\
      F_{12}^+(t,u) &=& F_{12}(t,u) + F_{12}(u,t)\,, \nonumber\\
      F_3^-(t,u) &=& F_3(t,u) - F_3(u,t)\,, \nonumber\\
      F_3^+(t,u) &=& F_3(t,u) + F_3(u,t)\,,
    \end{eqnarray}
leaving six physically relevant functions:
\begin{eqnarray}
	\label{eq::FFlincomb}
	F_{12}^+(t,u),\quad F_{12}^-(t,u),\quad F_2^-(t,u),\quad
    F_3^+(t,u),\quad F_3^-(t,u),\quad F_4(t,u),
\end{eqnarray}
where only $F_{12}^+(t,u)$ has contributions from triangle-type diagrams.
$F_k^-(t,u)$ (with $k=12,2,3$) and $F_4(t,u)$ are anti-symmetric w.r.t.\ the
exchange of the arguments $t$,$u$, and $F_k^+(t,u)$ (with $k=12,3$) are
symmetric.  At leading order $F_3^-(t,u) = 0$, however it has non-zero
contributions starting at NLO.

For the computation of the one- and two-loop Feynman diagrams (some examples
are shown in Fig.~\ref{fig::diags}) in the high-energy limit, we proceed as
follows: After producing the amplitude we perform a Taylor expansion in the
$Z$ and Higgs boson masses (since $m_Z^2,m_H^2 \ll m_t^2$), leaving one-
and two-loop integrals which depend only
on $s$, $t$ and $m_t$.  Using integration-by-parts reduction techniques with
the programs \texttt{FIRE~6}~\cite{Smirnov:2019qkx} and
\texttt{LiteRed}~\cite{Lee:2013mka}, these
integrals can be reduced to the same basis of 161 two-loop master integrals as in
Refs.~\cite{Davies:2018ood,Davies:2018qvx}, in which they were computed as an
expansion in the high-energy limit to order $m_t^{32}$. Inserting these
expansions into the amplitude yields its high-energy approximation.
We use \texttt{FORM~4.2}~\cite{Ruijl:2017dtg} for most stages of the
computation. The
calculation is performed in the covariant $R_\xi$ gauge and we allow for a
general electroweak gauge parameter $\xi_Z$ which appears in the propagators
of the $Z$ boson and Goldstone boson $\chi$.  Thus only the triangle-type
diagrams depend on $\xi_Z$, and this dependence cancels upon summing the $Z$
and $\chi$ contributions.

We also investigate the large-$m_t$ limit of the form factors. This expansion
is straightforward and proceeds in analogy
to~\cite{Davies:2018qvx,Davies:2020lpf}. The programs \texttt{q2e} and
\texttt{exp}~\cite{Harlander:1997zb,Seidensticker:1999bb} are used to produce
an asymptotic expansion for $m_t \gg q_1,q_2,q_3$, again performed using
\texttt{FORM}, yielding products of
massive vacuum integrals and massless three-point integrals. Results for the
$gg\to ZH$ amplitude, expanded to order $1/m_t^8$, have been previously
published in~\cite{Hasselhuhn:2016rqt}; here we provide one additional term in
this expansion, to order $1/m_t^{10}$, and provide results at the level of the
form factors.

The renormalization of the ultra-violet (UV) divergences proceeds in a similar way
as for the processes $gg\to HH$~\cite{Davies:2018qvx} and $gg\to ZZ$~\cite{Davies:2020lpf}.
In particular, we work in the six-flavour theory and renormalize the top quark mass on shell
and the strong coupling $\alpha_s$ in the $\overline{\rm MS}$ scheme.
In addition, our treatment of $\gamma_5$ (see Appendix~\ref{app::projg5} for more
details) requires that we apply additional finite
renormalization to the axial and pseudo-scalar currents~\cite{Larin:1993tq},
\begin{eqnarray*}
  Z_{5A} &=& 1 - \frac{\alpha_s}{\pi} C_F + {\cal O}(\epsilon)\,,\nonumber\\
  Z_{5P} &=& 1 - 2 \frac{\alpha_s}{\pi} C_F + {\cal O}(\epsilon)\,.
\end{eqnarray*}
The subtraction of the infra-red poles proceeds according to Ref.~\cite{Catani:1998bh},
using the conventions of
Refs.~\cite{Davies:2018qvx,Davies:2020lpf}.
The subtraction has the form~\cite{Catani:1998bh}
\begin{eqnarray}
  F^{(1)} &=& F^{(1),\rm IR} - K_g^{(1)} F^{(0)} \,\,=\,\,
              \tilde{F}^{(1)} + \beta_0
              \log\left(\frac{\mu^2}{-s-i\delta}\right) F^{(0)}\,,
  \label{eq::FIRsub}
\end{eqnarray}
with $\beta_0 = 11 C_A/12 - T n_f/3$ and $F^{(1),\rm IR}$ is UV renormalized
but still infra-red (IR) divergent. $F^{(1)}$ is finite.  An explicit
expression for $K_g$ is given in Eq.~(2.3) of Ref.~\cite{Davies:2018qvx}.
After the second equality sign in Eq.~(\ref{eq::FIRsub}) we make the $\mu$
dependence explicit. Note that below we only need $\tilde{F}^{(1)}$ to
construct the NLO cross-section.

In analogy to the process $gg\to HH$~\cite{Heinrich:2017kxx} we define the
NLO virtual finite cross-section for $gg\to ZH$ as
\begin{eqnarray}
  \tilde{{\cal V}}_{\rm fin} 
  &=& \frac{G_F^2 m_Z^2}{16s^2} \left(\frac{\alpha_s}{\pi}\right)^2
      \sum_{\lambda_1,\lambda_2,\lambda_3}
      \Bigg\{
      \left[ \tilde{A}_{\rm sub}^{\mu\nu\rho} \tilde{A}_{\rm
      sub}^{\star,\mu^\prime\nu^\prime\rho^\prime} \right]^{(1)}
      + \frac{C_A}{2}\left(\pi^2 - \log^2{\frac{\mu^2}{s}}\right) \left[ \tilde{A}_{\rm sub}^{\mu\nu\rho} \tilde{A}_{\rm
      sub}^{\star,\mu^\prime\nu^\prime\rho^\prime} \right]^{(0)} \Bigg\} 
      \nonumber\\&&\mbox{}\qquad \times 
      \varepsilon_{\lambda_1,\mu}(p_1)\:
      \varepsilon^\star_{\lambda_1,\mu^\prime}(p_1)\:
      \varepsilon_{\lambda_2,\nu}(p_2)\:
      \varepsilon^\star_{\lambda_2,\nu^\prime}(p_2)\:
      \varepsilon_{\lambda_3,\rho}(p_3)\:
      \varepsilon^\star_{\lambda_3,\rho^\prime}(p_3)\,,
      \label{eq::Vfin_ZH}
\end{eqnarray}
where the form factors entering the amplitude
$\tilde{A}_{\rm sub}^{\mu\nu\rho}$ are the IR-subtracted finite form factors
$\tilde{F}^{(1)}$ of Eq.~(\ref{eq::FIRsub}).  The superscripts ``(0)'' and
``(1)'' after the square brackets in Eq.~(\ref{eq::Vfin_ZH}) indicate that we
take the coefficients of $(\alpha_s/\pi)^0$ and $(\alpha_s/\pi)^1$,
respectively, of the squared amplitude, in accordance with
Eq.~(\ref{eq::FasExp}).  For the discussion in Section~\ref{sec::vfin} it is
convenient to introduce the $\alpha_s$-independent quantity
\begin{eqnarray}
  {\cal V}_{\rm fin} &=& \frac{\tilde{{\cal V}}_{\rm fin}}{\alpha_s^2}
                         \,.
                         \label{eq::Vfin_norm}
\end{eqnarray}


\section{Comparison at leading order}
\label{sec::1l-comparison}

In this section we compare our high-energy expansion with the exact LO result.
We implement this by using {\tt FeynArts}~\cite{Hahn:2000kx} to generate the
amplitude and {\tt FormCalc}~\cite{Hahn:1998yk} to compute it, performing a
Passarino-Veltman reduction to three- and four-point one-loop scalar
integrals.  Schouten identities allow us to write the result in terms of the
tensor structures and form factors of Eqs.~(\ref{eq::Amunuro}) and
(\ref{eq::F12_F-+}).  We use {\tt Package-X}~\cite{Patel:2016fam} to evaluate
the Passarino-Veltman functions with high precision and to produce an analytic
expansion in the limit of a large top quark mass.  We have verified that our
implementation of the exact LO result reproduces the cross-sections provided
in the
literature~\cite{Brein:2004ue,Altenkamp:2012sx,Hasselhuhn:2016rqt}. Additionally
we have compared the large-$m_t$ expansion derived from this result with a
direct expansion of the amplitude as described above.

\begin{figure}[t]
  \begin{center}
      \subcaptionbox{}{
         \includegraphics[width=.47\textwidth]{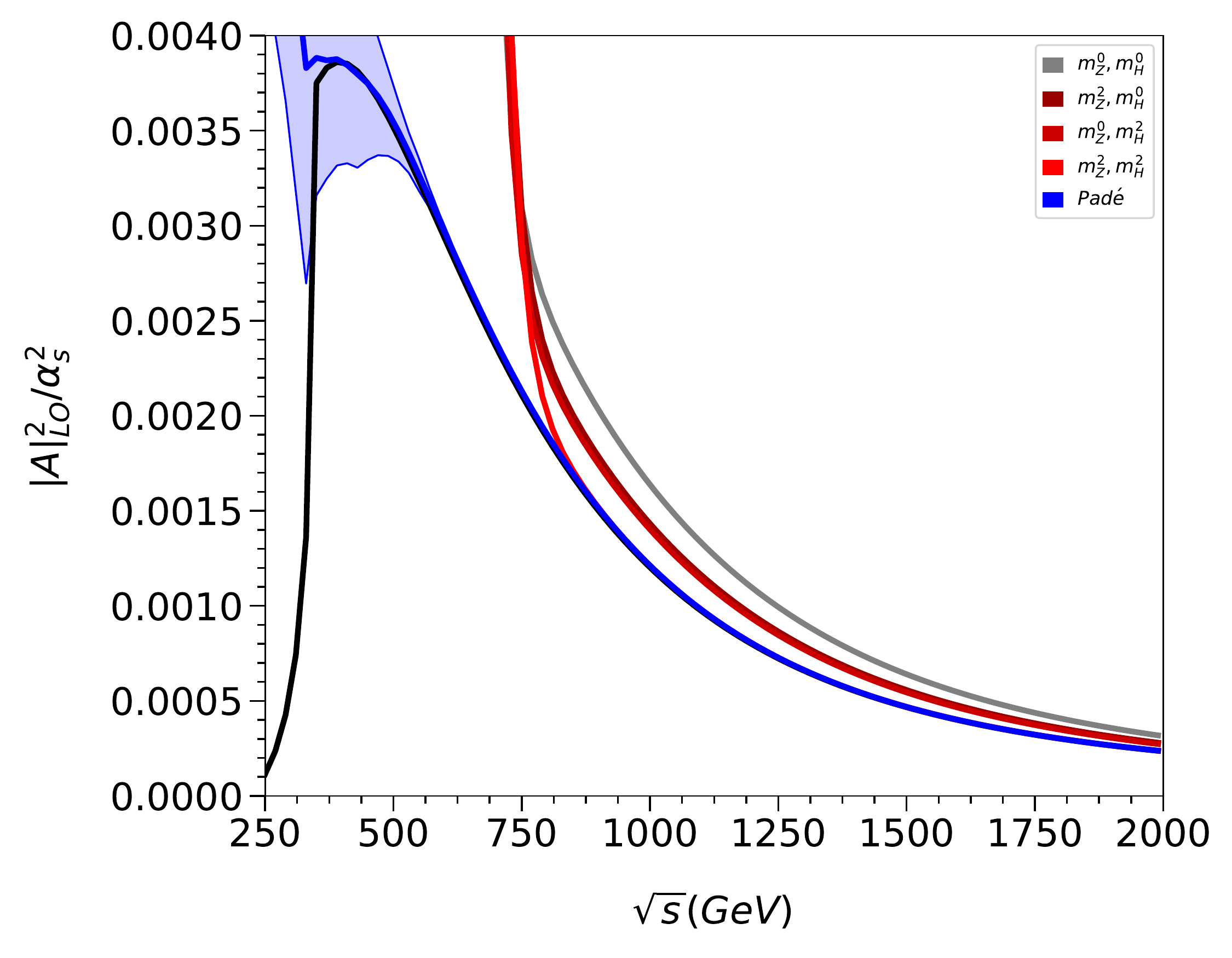}
      } \\
      \subcaptionbox{}{
         \includegraphics[width=.47\textwidth]{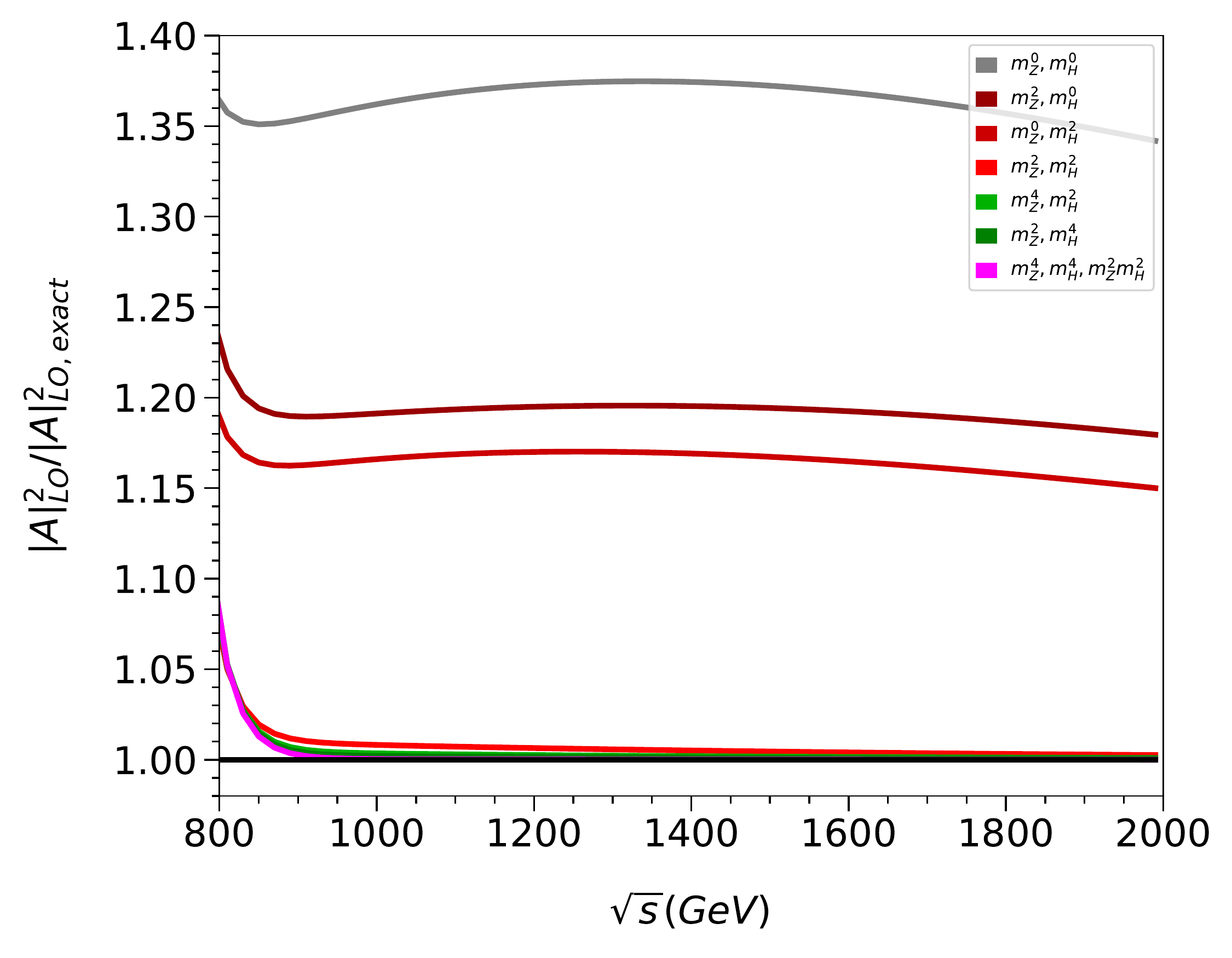}
      }
      \subcaptionbox{}{
         \includegraphics[width=.47\textwidth]{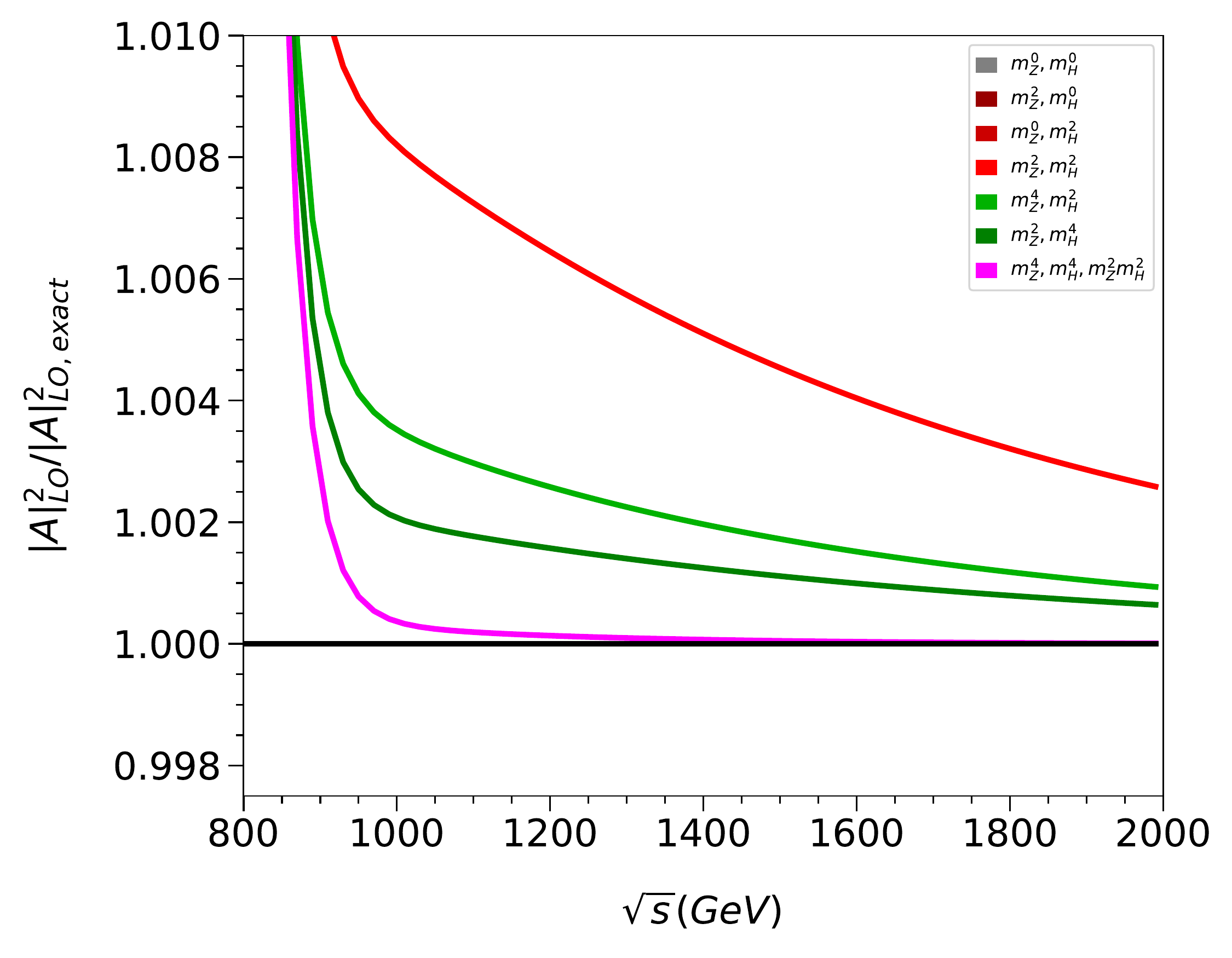}
      }
  \end{center}
  \caption{\label{fig::LO_amp2} (a) LO squared amplitude and (b) ratio of
    expansions to the
    exact result. Plot (c) is a zoomed-in version of (b). Note that for better
    readability in (a) no quartic corrections are shown; the black curve in (a)
    refers to the exact result and the blue curve (and associated uncertainty band)
    is the result obtained from Pad\'e approximation.}
\end{figure}

In Fig.~\ref{fig::LO_amp2}
we show the squared amplitude at LO, as a function of
$\sqrt{s}$, for a fixed scattering angle $\theta=\pi/2$. The solid black lines
correspond to the exact result. The coloured lines correspond to different expansion
depths in $m_Z$ and $m_H$, as detailed in the plot legend. All of the latter
include high-energy expansion terms up to $m_t^{32}$. 
The red curve demonstrates that after including quadratic terms both in $m_H$
and $m_Z$, the deviation from the exact result is below 1\% (for
$\sqrt{s}\gsim 1000$~GeV).  The $m_H^2$ corrections appear to be more
important than the $m_Z^2$ corrections.  The inclusion of the quartic terms
(shown as green and pink curves, best visible in
Fig.~\ref{fig::LO_amp2}(c)) improves
the accuracy of the expansion, leading to an almost negligible deviation from
the exact expression.  These terms are much harder to compute, however, so in the
NLO results we will restrict the approximation to the quadratic corrections
only.  The solid blue curve and associated uncertainty band in
Fig.~\ref{fig::LO_amp2}(a) shows the result of a procedure to improve the
expansions based on Pad\'e approximants, which is discussed in more detail in
Section~\ref{sec::highenergy}. Here it is based on the expansion to quadratic
order in $m_H$ and $m_Z$, confirming that our computation of the NLO amplitude
only to this order is sufficient.

In Fig.~\ref{fig::LO_amp2} and in the following sections, we use
the following parameter values:
\begin{eqnarray}
  m_Z = 91.1876~\mbox{GeV},\,
  m_t = 172.9~\mbox{GeV},\,
  m_H = 125.1~\mbox{GeV},\,
  G_F = 1.16638\!\times\!10^{-5} \mbox{GeV}^{-2}.
\end{eqnarray}


\section{NLO form factors: large-$m_t$ limit}
\label{sec::lme}

In this section we discuss the large-$m_t$ expansion of the form factors at
NLO.  While the expansion of $\tilde{F}_{12}^+$ starts at $m_t^0$, we find that the
other form factors of Eq.~(\ref{eq::FFlincomb}) exhibit some cancellation in
the leading contributions. In particular, $\tilde{F}_{12}^-$, $\tilde{F}_2^-$, $\tilde{F}_3^+$ and
$\tilde{F}_4$ start only at $1/m_t^4$, and $\tilde{F}_3^-$ starts at $1/m_t^6$. 

We now present the leading terms of the large-$m_t$ expansion to establish our notation.
For brevity, we restrict ourselves to $\tilde{F}_{12}^+$. Expressions for the expansion of all form
factors to $1/m_t^{10}$ are provided in the ancillary files of this paper~\cite{progdata}.
Our result reads
\begin{align}
\tilde{F}_{12}^{+,(1)} &=
-2C_A+
\frac{s}{m_t^2}\, \left(\frac{11 C_A}{72}  - \frac{C_F}{4}\right)
+ \frac{1}{m_t^4} \left(C_F\,s\left[\frac{222 m_H^2 + 162 m_Z^2 - 263 s}{2160}\right]\right.
\nonumber\\&
+ C_A \left( \frac{1}{129600}\left[-75\left\{m_H^4 + m_Z^4 - \frac{6}{5} m_H^2 m_Z^2\right\} 
+ 4011 m_H^2 s + 6713 m_Z^2 s \right.\right.
\nonumber\\&
\left.\left.\vphantom{\frac{1}{1}}
- 2646 s^2 + 60 \left\{m_H^2 + m_Z^2 -s- t\right\} t \right]
- s\,\log{\left(\frac{-s}{m_t^2}\right)}
\left[\frac{9 m_H^2 + 26 m_Z^2 - 9 s}{2160}\right] \right)\,,
\end{align}
where $C_A=3$ and $C_F=4/3$ are the quadratic Casimir invariants of $SU(3)$.

We have verified that after constructing $\tilde{{\cal V}}_{\rm fin}$ in
Eq.~(\ref{eq::Vfin_ZH}) we find agreement with the results of
Ref.~\cite{Hasselhuhn:2016rqt} up to order $1/m_t^8$.


\section{\label{sec::highenergy}NLO form factors: high-energy limit}

\begin{figure}
  \begin{tabular}{cc}
    \hspace{-5mm} \includegraphics[width=0.51\linewidth]{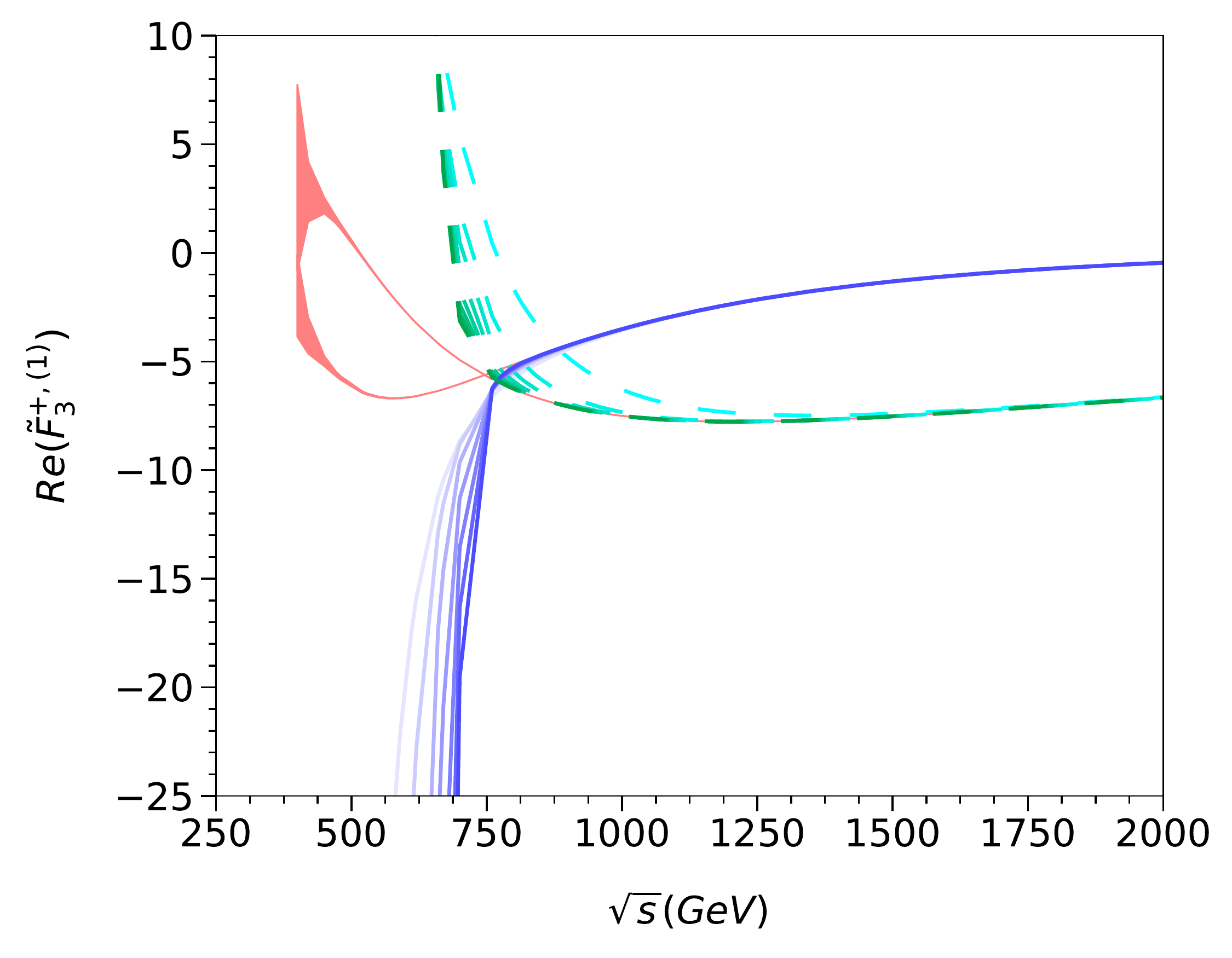} \hspace{-6mm}
    & \includegraphics[width=0.51\linewidth]{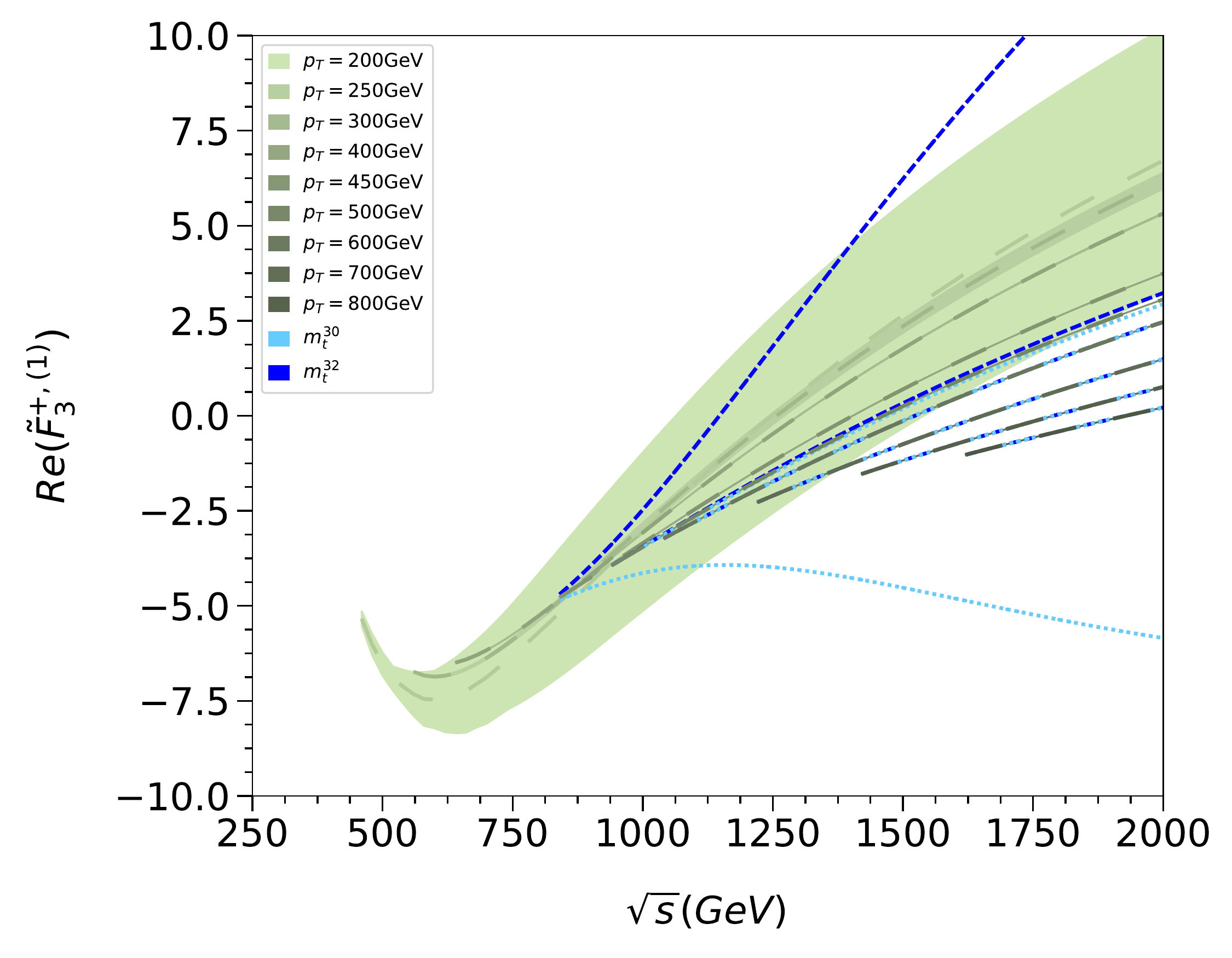}
    \\
    (a) & (b)
  \end{tabular}
  \caption{\label{fig::FF-HE-pi2} Results for $\tilde{F}_{3}^+$ as a function of
    $\sqrt{s}$ for fixed $\theta=\pi/2$ (a) and for fixed $p_T$ (b).  In (a)
    dashed and solid lines correspond to the imaginary and real parts,
    respectively. The red curves in (a) represent the Pad\'e
    results. In (b) only the real part is shown, and the Pad\'e results are shown
    as green dashed lines. The high-energy expansions up to order $m_t^{30}$ and
    $m_t^{32}$ are shown in blue. The widely-separated pair of curves correspond
    to $p_T=400$~GeV.
}
\end{figure}

We now turn to the high-energy expansion of the NLO form factors.  The
analytic expressions are large, so we refrain from showing them
here but we provide analytic expressions for all form factors in the
ancillary files of this paper~\cite{progdata}.  For illustration we discuss in
the following the results for $\tilde{F}_{3}^+$.

In Fig.~\ref{fig::FF-HE-pi2}(a) we plot $\tilde{F}_{3}^+$ as a function of $\sqrt{s}$.
We can see that the high-energy expansions of both the real (blue solid
lines) and imaginary parts (green dashed lines) converge well for
$\sqrt{s} \gsim 800$~GeV, which is in analogy to
$gg\to HH$~\cite{Davies:2018qvx} and $gg\to ZZ$~\cite{Davies:2020lpf}.  Note
that the lighter-coloured curves include fewer $m_t$ expansion terms; the
darkest lines show the expansion up to $m_t^{32}$.

As in previous publications on $gg\to HH$ and $gg\to ZZ$, we make use of
Pad\'e approximants to improve the description of the high-energy
expansion. The methodology used follows that of Section~4 of
Ref.~\cite{Davies:2020lpf} and so is not described in detail, but is only
summarized briefly here. The expansion is used to produce 28
different Pad\'e approximants, which are combined to produce a central value
and error estimate for the approximation.  In Fig.~\ref{fig::FF-HE-pi2}(a) the
Pad\'e results are shown as red bands. The width of the bands denote
the uncertainty estimates.  For regions in which the high-energy
expansion converges, the Pad\'e-based approximation reproduces the
expansion. However, it also produces reliable results for smaller values of
$\sqrt{s}$, as can be expected from the comparison with the LO result shown in
Fig.~\ref{fig::LO_amp2}(a).

In Fig.~\ref{fig::FF-HE-pi2}(b) we show $\tilde{F}_{3}^+$ for the fixed values of
$p_T = 200, \ldots, 800$~GeV. The blue (dashed and dotted) curves correspond to the
high-energy expansions and the green (dashed) curves to the Pad\'e
results. For all Pad\'e curves we also show the corresponding uncertainty band, which
for $p_T=200$~GeV is relatively large but for $p_T=250$~GeV the
uncertainty band is already quite small; it is completely negligible for
higher values of $p_T$.  Note that the high-energy expansions are only shown for
$p_T\ge 400$~GeV; for lower $p_T$ values the curves lie
far outside of the plot range. For $p_T\gsim 450$~GeV the expansions converge
and are very close to the Pad\'e results.
For $p_T=400$~GeV, while the expansions initially
agree with each other and the Pad\'e close to $\sqrt{s}=800$~GeV, they diverge
for larger values of $\sqrt{s}$. We recall here that the high-energy expansion
is an expansion in $m_t^2/s$, $m_t^2/t$ and $m_t^2/u$. For a fixed value of $p_T$,
increasing $\sqrt{s}$ can lead to values of $t$ or $u$ which are not large enough
for convergence.

In Section~\ref{sec::vfin} we will apply the Pad\'e procedure to the virtual
finite cross-section, in order to compare our results with a
state-of-the-art numerical evaluation at NLO~\cite{Heinrich_etal}.


\section{\label{sec::vfin}NLO virtual finite cross-section}

\begin{figure}[t]
  \includegraphics[width=\textwidth]{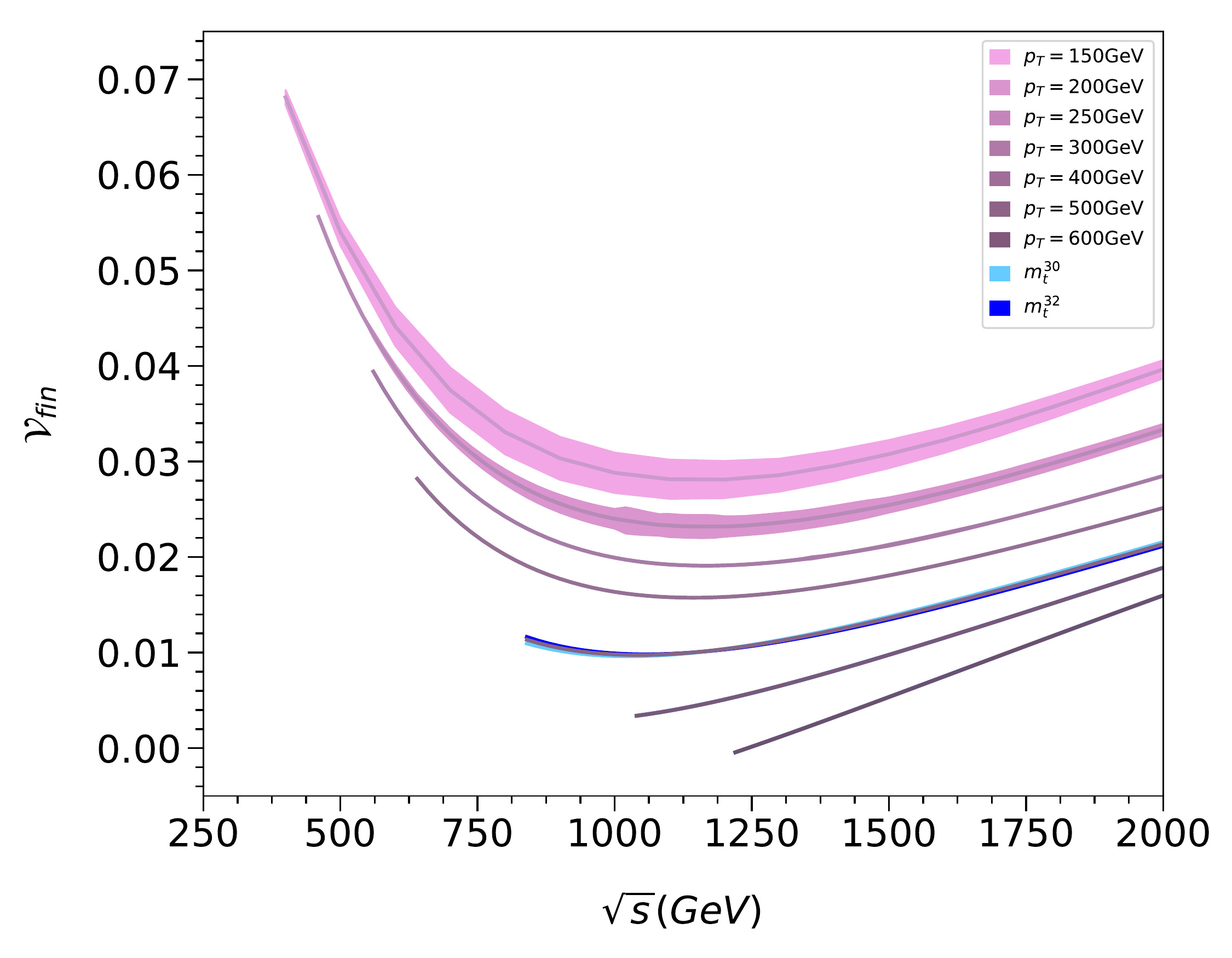}
  \caption{\label{fig::Vfin}${\cal V}_{\rm fin}$ as a function of $\sqrt{s}$
    for eight values of $p_T$, at the scale $\mu^2=s$. The uncertainty estimate
    of the Pad\'e procedure is displayed as light-coloured bands.
    For $p_T \geq 350$~GeV,
    we also show two high-energy expansion curves including terms up to order
    $m_t^{30}$ and $m_t^{32}$.
     }
\end{figure}

Our starting point is $\tilde{{\cal V}}_{\rm fin}$ in Eq.~(\ref{eq::Vfin_ZH}).
For the LO form factors we use the exact results and for the two-loop form
factors we use the high-energy expansion. This allows us to write
${\cal V}_{\rm fin}$ in Eq.~(\ref{eq::Vfin_norm}) as an expansion in $m_t$ up
to order $m_t^{32}$.\footnote{Note the double-triangle contribution, which is
  known analytically (see Appendix~\ref{app::1PR} and
  Ref.~\cite{Hasselhuhn:2016rqt}), is not included in our numerical results for
  ${\cal V}_{\rm fin}$.}  At this point we can apply the Pad\'e approximation
procedure as described in Section~\ref{sec::highenergy}.  In
Fig.~\ref{fig::Vfin} we show ${\cal V}_{\rm fin}$ for several fixed values of
$p_T$ as a function of $\sqrt{s}$. For $p_T=400$~GeV and larger we also show
two high-energy expansion curves, which include terms up to $m_t^{30}$ and
$m_t^{32}$. These curves agree well with each other and with the Pad\'e
approximation which they produce.  For lower values of $p_T$, these curves do
not agree with each other, and are not visible within the plot range.

For $p_T=150$~GeV Pad\'e procedure produces stable results with an uncertainty
of about 10\%. For $p_T=200$~GeV the uncertainties are notably smaller and
for higher $p_T$ values they are negligible.  We advocate to use results based
on our approach for $p_T\gsim 200$~GeV.  For $p_T \approx 150$~GeV the Pad\'e
approach provides important results for cross-checks. For lower values of
$p_T$ other methods have to be used for the calculation of $gg\to ZH$, see
Ref.~\cite{Heinrich_etal}.

We have compared our results with those of Ref.~\cite{Heinrich_etal},
which are obtained numerically, but without making any expansions.  In the
high-energy region, for 104 kinematic points with $p_T\geq 200$~GeV we agree
to within 2-3\%, and for 42 points with $150 \leq p_T < 200$~GeV we agree to
within 10\% and our values are consistent to within the uncertainty of the
Pad\'e procedure.  We also construct ${\cal V}_{\rm fin}$ using the
large-$m_t$ expanded NLO form factors of Section~\ref{sec::lme} and find
that for 120 points with $\sqrt{s}\leq 284$~GeV, we agree to within 1\%. For
larger values of $\sqrt{s}$, as expected, the large-$m_t$ expansion starts
to diverge significantly from the numerical results as one approaches the
top quark threshold at $\sqrt{s}\approx 346$~GeV.


\section{Conclusions}
\label{sec::concl}

In this paper we consider two-loop NLO corrections to the Higgs Strahlung
process $gg\to ZH$. The corresponding Feynman integrals involve five
dimensionful parameters ($s$, $t$, $m_t$, $m_Z$ and $m_H$) which makes an
analytic calculation impossible with the current state-of-the-art techniques.
Numerical computations are also very challenging, however they have recently
been achieved in Ref.~\cite{Heinrich_etal}.
In Section~\ref{sec::vfin} we have discussed the cross-check
of our results against these.

In our approach we use the hierarchy between the top quark, Higgs and $Z$
boson masses and perform an expansion for $m_t^2 \gg m_Z^2,m_H^2$,
effectively eliminating the dependence on $m_Z$ and $m_H$ from the
integrals. We show at one-loop order that the expansion converges quickly,
which allows us to truncate the expansion at NLO after the quadratic terms.
As far as the remaining scales are concerned, we concentrate on the
high-energy region where $s,t \gg m_t^2$. We expand our master integrals in
this limit such that we obtain results for the form factors including
expansion terms up to $m_t^{32}$. This allows us to construct, for each
phase-space point (e.g., a particular pair of $\sqrt{s}$, $p_T$ values)
a set of Pad\'e approximants, which considerably extend the region of
convergence of our expansion. Our approach provides both central values and
uncertainty estimates.

We provide results for all form factors involved in the $gg\to ZH$ amplitude,
both at one- and two-loop order.  Furthermore, we provide relations between
the form factors and helicity amplitudes in Appendix~\ref{app::helamp}.  The
main emphasis of the paper is the IR subtracted virtual corrections to the partonic
cross-section, which can be combined with other (e.g. Drell-Yan--like)
corrections to $gg\to ZH$.  Our method provides precise results for
$p_T\gsim 200$~GeV with almost negligible uncertainties.  For lower values of
$p_T$ our uncertainties increase, in particular for smaller values of
$\sqrt{s}$.  The analytic results for the form factors obtained in this paper
can be obtained in electronic form from Ref.~\cite{progdata}.


\smallskip

{\bf Acknowledgements.}  We thank the authors of Ref.~\cite{Heinrich_etal} for
making their results available for comparison, prior to publication.
The work of J.D. was in part supported by the Science and Technology Research
Council (STFC) under the Consolidated Grant ST/T00102X/1.
The work of
G.M. was in part supported by JSPS KAKENHI (No. JP20J00328).  This research
was supported by the Deutsche Forschungsgemeinschaft (DFG, German Research
Foundation) under grant 396021762 --- TRR 257 ``Particle Physics Phenomenology
after the Higgs Discovery''.


\appendix


\section{\label{app::1PR}Double-triangle contribution}

In this appendix we present results for the one-particle reducible
double-triangle contribution, as shown in the first diagram in the second row of
Fig.~\ref{fig::diags}. We have computed the six form factors in
Eq.~(\ref{eq::FFlincomb}). The results can be expressed in terms of the
functions (see also Ref.~\cite{Degrassi:2016vss} where the corresponding
contributions for $gg\to HH$ were computed)
\begin{eqnarray}
  {B_0(x,0,0)} &=& 2 - \log\frac{-x}{\mu^2}
                 \,,\nonumber\\
  {B_0(x,m^2,m^2)} &=& 2 + \beta_x \log\frac{\beta_x-1}{\beta_x+1} - \log\frac{m^2}{\mu^2}
                 \,,\nonumber\\
  {C_0(x,y,0,m^2,m^2,m^2)} &=&  \frac{1}{2(x-y)}\left(\log^2\frac{\beta_x+1}{\beta_x-1} - \log^2\frac{\beta_y+1}{\beta_y-1}\right)\,,
\end{eqnarray}
with $\beta_x = \sqrt{1-4m^2/x}$ and $\beta_y = \sqrt{1-4m^2/y}$.
Our explicit calculation shows that $F_3^+(t,u)=F_3^-(t,u)=F_4(t,u)=0$. We
furthermore find that $F_{12}(t,u)$ is equal to $F_2(t,u)$ with
\begin{align}
 F_2(t,u)=&
    \frac{8 m_t^2 m_Z^2 s}{\left(m_H^2-u\right)^2 \left(m_Z^2-u\right)^2}
    \bigg\{2 u \left[ B_0(m_H^2,m_t^2,m_t^2)- B_0(u,m_t^2,m_t^2)\right]
    \nonumber\\
  & 
    + \left(m_H^2-u\right) \left[\left(-m_H^2+4 m_t^2+u\right)
    C_0(u,m_H^2,0,m_t^2,m_t^2,m_t^2)+2\right] \bigg\}
    \nonumber\\
  & 
    \quad 
    \times \bigg[ B_0(u,m_t^2,m_t^2)-
    B_0(m_Z^2,m_t^2,m_t^2)
    + B_0(m_Z^2,0,0) - B_0(u,0,0)
    \nonumber\\
  & 
    \quad\quad
    + 2m_t^2 \frac{u-m_Z^2}{m_Z^2}
    C_0(u,m_Z^2,0,m_t^2,m_t^2,m_t^2)
    \bigg]
    \,,
    \label{eq::FF_1PR}
\end{align}
which can be used to construct $F_{12}^+(t,u)$, $F_{12}^-(t,u)$ and
$F_{2}^-(t,u)$ (cf. Section~\ref{sec::notation}).
We note that it is straightforward to expand the results in
Eq.~(\ref{eq::FF_1PR}) in the large- and small-$m_t$ limit.

The contribution of the double-triangle diagrams to the squared matrix element
have been computed in Ref.~\cite{Hasselhuhn:2016rqt} and implemented in the
corresponding computer program, see Appendix of
Ref.~\cite{Hasselhuhn:2016rqt}.


\section{\label{app::projg5}Projectors and $\gamma_5$}

Each LO and NLO diagram contributing to the $gg\to ZH$ amplitude contains a quark
with either an axial-vector coupling to a $Z$ boson, or a pseudo-scalar coupling
to a Goldstone boson. The $\gamma_5$ matrix present in these couplings is not defined
in the $d=4-2\epsilon$ space-time dimensions of dimensional regularization.
The couplings are re-written in terms of anti-symmetric tensors, according to
\cite{Larin:1993tq}, as
\begin{eqnarray}
  \gamma^\mu \gamma^5 &=& 
  \frac{i}{12} \, \varepsilon^{\mu\nu\rho\sigma} 
  \left( 
    \gamma_{\nu}\gamma_\rho\gamma_{\sigma} - 
    \gamma_{\sigma}\gamma_\rho\gamma_{\nu}
  \right)
  \,,
  \nonumber\\
  \gamma^5 &=& 
  \frac{i}{96} \, \varepsilon^{\mu\nu\rho\sigma} 
  \left( 
    \gamma_{\mu}\gamma_{\nu}\gamma_\rho\gamma_{\sigma} +
    \gamma_{\sigma}\gamma_{\rho}\gamma_\nu\gamma_{\mu} -
    \gamma_{\nu}\gamma_\rho\gamma_{\sigma}\gamma_{\mu} -
    \gamma_{\mu}\gamma_\sigma\gamma_\rho\gamma_\nu
  \right)
  \,,
  \label{eq::gamma5}
\end{eqnarray}
which allow one to compute the loop integrals in two ways.
The first is to ensure the $\varepsilon$ tensors are not contracted until one
can safely work in four dimensions, by solving tensor loop integrals, performing
UV renormalization and IR subtraction (as detailed in Eq.~(\ref{eq::FIRsub})) and
then finally contracting with the $\varepsilon$ tensors.

An alternative approach, which avoids the need to compute tensor integrals, is to
project the amplitude onto the 60 possible rank-three Lorentz structures which can be
formed from the three independent momenta and the $\varepsilon$ tensor. Since each
term of the projectors onto these structures contains an $\varepsilon$ tensor,
the result of their contraction with the r.h.s.\ of Eq.~(\ref{eq::gamma5}) can be
treated correctly in $d$ dimensions.
The 60 projectors act on the amplitude as
\begin{eqnarray}
	P_i^{\mu_1 \mu_2 \mu_3}\: A_{\mu_1 \mu_2 \mu_3} = F_i
\end{eqnarray}
to produce 60 form factors $F_i$, which can be reduced to a minimal set (see the
discussion around Eq.~(\ref{eq::Amunuro})).
Each projector $P_i^{\mu_1 \mu_2 \mu_3}$ can be written generically as
\begin{eqnarray}
	  C_{i,1} \varepsilon^{\mu_1 \mu_2 \mu_3 \nu_1} q_{1\, \nu_1}
	+ C_{i,2} \varepsilon^{\mu_1 \mu_2 \mu_3 \nu_1} q_{2\, \nu_1}
	+ \cdots
	+ C_{i,60} g^{\mu_1 \mu_2} \varepsilon^{\mu_3 \nu_1 \nu_2 \nu_3}
		q_{1\, \nu_1} q_{2\, \nu_2} q_{3\, \nu_3}\,,
\end{eqnarray}
and we contract this general form with our amplitude. In the final result,
we specify each of the 60 sets of coefficients $\{C_{i,j}\}$ in order to obtain
the 60 form factors $F_i$.


\section{\label{app::helamp}Helicity amplitudes}

In this appendix we describe how one can obtain the helicity amplitudes for
the process $gg\to ZH$ from the tensor decomposition which we have introduced
in Section~\ref{sec::notation}. In analogy to Eq.~(\ref{eq::helicity_amp}) we introduce
\begin{eqnarray}
  \tilde{\mathcal{M}}_{\lambda_1,\lambda_2,\lambda_3} &=& \tilde{A}^{\mu\nu\rho}
               \varepsilon_{\lambda_1,\mu}(p_1)\varepsilon_{\lambda_2,\nu}(p_2)
               \varepsilon_{\lambda_3,\rho}(p_3)
               \,.
\end{eqnarray}
We furthermore specify the (contravariant) external momenta and polarization
vectors as follows:
\begin{align}
&
p_1=\frac{\sqrt{s}}{2}
\left(
\begin{array}{c}
1\\ 0\\ 0\\ 1
\end{array}
\right)
\!,\,\,
p_2=\frac{\sqrt{s}}{2}
\left(
\begin{array}{c}
1\\ 0\\ 0\\ -1
\end{array}
\right)
\!,\,\,
p_3=\frac{\sqrt{s}}{2}
\left(
\begin{array}{c}
-y_1\\
0\\ -y_2\sin \theta \\ y_2\cos \theta 
\end{array}
\right)
\!,\,\,
p_4=\frac{\sqrt{s}}{2}
\left(
\begin{array}{c}
-2+y_1\\
0\\  y_2\sin \theta \\ -y_2\cos \theta 
\end{array}
\right)
,
\nonumber\\
&
\varepsilon_+ (p_1)
=
[ \varepsilon_- (p_1) ]^\star
=
\frac{1}{\sqrt{2}}
\left(
\begin{array}{c}
0\\ i\\ 1\\ 0
\end{array}
\right)
\!,\,\,
\qquad
\varepsilon_+ (p_2)
=
[ \varepsilon_- (p_2) ]^\star
=
\frac{1}{\sqrt{2}}
\left(
\begin{array}{c}
0\\ -i\\ 1\\ 0
\end{array}
\right)
\!,\,\,
\nonumber\\
&
\varepsilon_+ (p_3)
=
[ \varepsilon_- (p_3) ]^\star
=
\frac{1}{\sqrt{2}}
\left(
\begin{array}{c}
0\\ i\\ \cos \theta \\ \sin \theta
\end{array}
\right)
\,,
\quad
\varepsilon_0 (p_3)
=
\frac{\sqrt{s}}{2m_Z}
\left(
\begin{array}{c}
 y_2\\
  0\\
 y_1 \sin \theta \\
   -y_1 \cos \theta
\end{array}
\right),
\nonumber\\
&y_{1}=1+\frac{m_{Z}^{2}-m_{H}^{2}}{s}\,,\qquad
y_{2}=\beta=\sqrt{1-2 \frac{m_{Z}^{2}+m_{H}^{2}}{s}+\frac{\left(m_{Z}^{2}-m_{H}^{2}\right)^{2}}{s^{2}}}\,,
\label{eq:def-vectors}
\end{align}
where $\varepsilon_0$ denotes the
longitudinal components of polarization vectors.
Recall that all external momenta are defined as incoming,
see Eq.~(\ref{eq::mandelstam}).
We have chosen the convention for the polarisation vector of $p_2$,
following Ref.~\cite{Jacob:1959at},
such that $\varepsilon_+ (p_2)\to \varepsilon_- (p_1)$
in the center-of-momentum frame.
Furthermore, the polarization vectors satisfy
\begin{align}
\sum_{\lambda_{1}} 
\varepsilon_{\lambda_{1}, \mu}\left(p_{1}\right) 
\varepsilon_{\lambda_{1}, \mu^{\prime}}^{\star}\left(p_{1}\right)
&=-g_{\mu \mu^{\prime}}
+\frac{p_{1, \mu} p_{2, \mu^{\prime}}+p_{2, \mu} p_{1, \mu^{\prime}}}{p_{1} \cdot p_{2}}
\,,
\nonumber\\
\sum_{\lambda_{2}} 
\varepsilon_{\lambda_{2}, \nu}\left(p_{2}\right) 
\varepsilon_{\lambda_{2}, \nu^{\prime}}^{\star}\left(p_{2}\right)
&=-g_{\nu \nu^{\prime}}
+\frac{p_{2, \nu} p_{1, \nu^{\prime}}+p_{1, \nu} p_{2, \nu^{\prime}}}{p_{1} \cdot p_{2}}
\,,
\nonumber\\
\sum_{\lambda_{3}} 
\varepsilon_{\lambda_{3}, \rho}\left(p_{3}\right) 
\varepsilon_{\lambda_{3}, \rho^{\prime}}^{\star}\left(p_{3}\right)
&=-g_{\rho \rho^{\prime}}
+\frac{p_{3, \rho} p_{3, \rho^{\prime}}}{m_{Z}^{2}}
\,,
\end{align}
which means that we have fixed the gauge
for the external gauge bosons.
With the above choice,
some of the tensor structures in
$A_{a b}^{\mu \nu \rho}\left(p_{1}, p_{2}, p_{3}\right)$,
which are proportional to
either $p_1^\nu$ or $p_2^\mu$,
are irrelevant because
\begin{align}
\varepsilon_{\lambda_{1}}\left(p_{1}\right) \cdot p_2
=
\varepsilon_{\lambda_{2}}\left(p_{2}\right) \cdot p_1
=0
\,.
\end{align}
This reduces the number of Lorentz structures in Eq.~(\ref{eq::Amunuro})
from 14 to 8. We note that, as we will see in Eq.~(\ref{eq::hel_amps}),
the dependence on $F_2^+$ drops out in the helicity amplitude.

In total there are $2\times 2\times 3=12$ helicity amplitudes. However, due
to various symmetries, only 4 of them are independent.
With explicit calculation we find 
\begin{align}
&  \tilde{\mathcal{M}}_{-\lambda_1,-\lambda_2,-\lambda_3} =
-\tilde{\mathcal{M}}_{\lambda_1,\lambda_2,\lambda_3}   \,,\nonumber\\
&  \tilde{\mathcal{M}}_{++-} = -\tilde{\mathcal{M}}_{+++}
  \Big| _{\theta\to\theta+\pi, y_2\to-y_2}
  \,,\nonumber\\
 & \tilde{\mathcal{M}}_{+--} = -\tilde{\mathcal{M}}_{+-+}
  \Big| _{\theta\to\theta+\pi, y_2\to-y_2}
  \,,
  \label{eq:sym1}
\end{align}
yielding the 4 independent helicity amplitudes.
Note that the Mandelstam variables are invariant under the simultaneous replacements 
$\theta\to\theta+\pi, y_2\to-y_2$
and thus the form factors do not change.

For the evaluation of the $\varepsilon$ tensor,
we use the convention that
\begin{align}
\varepsilon^{0123}=+1,
\end{align}
and so some of the relevant contractions are as follows:
\begin{align}
\varepsilon ^{\mu\rho\alpha\beta}
\varepsilon_{\mu +} (p_1)
\varepsilon_{\rho +} (p_3)
p_{1\alpha}
p_{2\beta}
&=-
i\frac{s}{4}(1-\cos\theta)\,,
\nonumber
\\
\varepsilon ^{\mu\rho\alpha\beta}
\varepsilon_{\mu +} (p_1)
\varepsilon_{\rho +} (p_3)
p_{1\alpha}
p_{3\beta}
&=
i\frac{s}{8}(1-\cos\theta)(y_1-y_2)\,,
\nonumber
\\
\varepsilon ^{\rho\alpha\beta\gamma}
\varepsilon_{\rho +} (p_3)
p_{1\alpha}
p_{2\beta}
p_{3\gamma}
&=
-i\frac{s\sqrt{s}}{4\sqrt{2}} y_2 \sin\theta\,,
\nonumber
\\
\varepsilon ^{\mu\rho\alpha\beta}
\varepsilon_{\mu +} (p_1)
\varepsilon_{\rho 0} (p_3)
p_{1\alpha}
p_{2\beta}
&=
i\frac{s\sqrt{s}}{4\sqrt{2}m_Z} y_1 \sin\theta\,,
\nonumber
\\
\varepsilon ^{\mu\rho\alpha\beta}
\varepsilon_{\mu +} (p_1)
\varepsilon_{\rho 0} (p_3)
p_{1\alpha}
p_{3\beta}
&=
-i\frac{m_Z\sqrt{s}}{2\sqrt{2}} \sin\theta\,,
\nonumber
\\
\varepsilon ^{\mu\nu\rho\alpha}
\varepsilon_{\mu +} (p_1)
\varepsilon_{\nu +} (p_2)
\varepsilon_{\rho +} (p_3)
p_{1\alpha}
&=
i\frac{\sqrt{s}}{2\sqrt{2}} \sin\theta\,,
\nonumber
\\
\varepsilon ^{\mu\nu\rho\alpha}
\varepsilon_{\mu +} (p_1)
\varepsilon_{\nu +} (p_2)
\varepsilon_{\rho 0} (p_3)
p_{1\alpha}
&=
-i\frac{s}{4m_Z} (y_1\cos\theta +y_2)
\,.
\end{align}
In terms of the form factors, the 4 independent helicity amplitudes read
\begin{align}
\tilde{\mathcal{M}}_{+-0}
&=
-i\frac{s}{16m_Z} y_2 \sin ^2\theta (s y_1 F_2^- -2m_Z^2 F_3^-)
\nonumber\\
\tilde{\mathcal{M}}_{+-+}
&=i
\frac{s\sqrt{s}}{16\sqrt{2}} y_2 \sin \theta (1-\cos\theta) (2F_2^- -y_1 F_3^- - y_2 F_3^+ )
\nonumber\\
\tilde{\mathcal{M}}_{+++}
&=
i\frac{s\sqrt{s}}{16\sqrt{2}} \sin \theta
\left[
4F_{12}^-
-(y_1-y_2)(2F_2^-+4F_4+y_2F_3^-+y_2\cos\theta F_3^+)
\right]
\nonumber\\
\tilde{\mathcal{M}}_{++0}
&=
-i\frac{s}{8m_Z} 
\left[
(sy_1F_{12}^--2m_Z^2F_2^--4m_Z^2F_4)\cos\theta
-sy_2F_{12}^+
+m_Z^2y_2\sin^2\theta F_3^+
\right]
\label{eq::hel_amps}
\,,
\end{align}
where we have omitted the arguments of the form factors, $F(t,u)$.  

These helicity amplitudes have some general properties which are valid at any
loop order. It is useful to introduce the partial wave decomposition of the
amplitude~\cite{Jacob:1959at}
\begin{align}
\mathcal{M}_{\lambda_a,\lambda_b,\lambda_c}
=\frac{1}{4\pi}\sum_J (2J+1)
\bra{\lambda_c}S^J\ket{\lambda_a\lambda_b}
d^J_{\lambda_a-\lambda_b,\lambda_c} (\theta)
\label{eq::partial-wave}
\,,
\end{align}
where $\lambda_a,\lambda_b$ ($=\pm 1$ in this case) are the helicities of
initial state particles and $\lambda_c$ $(=\pm 1,0)$ is the helicity of the
$Z$ boson. $J$ is the total angular momentum of the system, $S^J$ is the
$J$-component of the S-matrix, and $d^J_{M,M'} (\theta)$ is the Wigner
small-$d$ function (see, e.g.~\cite{Sakurai:2011zz}).  In the case of
$\mathcal{M}_{++0}$,
\begin{align}
d^J_{\lambda_a-\lambda_b,\lambda_c} (\theta)
=d^J_{0,0}(\theta)
=P_J(\cos \theta)
\,,
\end{align}
where $P_J(x)$ are the Legendre polynomials, which are even functions for even
$J$, and odd functions for odd $J$.  Taking into account that
$F_{12}^-, F_2^-, F_3^-$ and $F_4$ are anti-symmetric in $t\leftrightarrow u$
exchange and thus odd functions in $\cos\theta$, whereas $F_{12}^+$ and
$F_3^+$ are symmetric in $t\leftrightarrow u$ exchange and thus even functions
in $\cos\theta$, we find that $\mathcal{M}_{++0}$ is an even function of
$\cos\theta$.  Using Eq.~\eqref{eq::partial-wave} and the property of the
Legendre polynomials mentioned above, we conclude that only $J$-even
components components contribute to $\mathcal{M}_{++0}$ and thus this
amplitude is a $J$-even channel.  In a similar discussion, it is also
straightforward to show that $\mathcal{M}_{+-0}$, expanded in terms of
$d^J_{2,0}(\theta)=P_J^2(\cos \theta)$ where $P_J^n$ is the associated
Legendre polynomial, is a $J$-odd channel, and that $\mathcal{M}_{+++}$,
expanded in terms of $d^J_{0,1}(\theta)=P_J^1(\cos \theta)$, is a $J$-even
channel.  On the other hand, $\mathcal{M}_{+-+}$ does not have such a feature.
In total, the squared amplitude should be symmetric in $t\leftrightarrow u$
exchange due to the fact that the two initial state gluons are
indistinguishable, and this symmetry is made apparent when summing all of the
squared helicity amplitudes.

The contribution from the triangle diagrams is present only in
$\mathcal{M}_{++0}$ via $F_{12}^+$, and this can be understood in the
following way.  Due to the Landau-Yang theorem, the on-shell mode of the
mediating $Z$-boson is forbidden; only the off-shell mode and the Goldstone
boson propagate.  Since the off-shell mode and the Goldstone boson behave as
scalars under rotation, they do not appear in $\mathcal{M}_{+-0}$ ($J$-odd) or
$\mathcal{M}_{+-+}$ ($J$-indefinite).  Furthermore, using the transversal
condition of the $Z$-boson, $\varepsilon_\pm (p_3) \cdot p_3=0$ one can also
derive that $\varepsilon_\pm (p_3) \cdot p_4=0$ and thus the transverse
components of the final state $Z$-boson do not couple to the
scalar-scalar-vector interaction (see, e.g., the explicit form of the Feynman
rule for the Goldstone-Higgs-$Z$ boson vertex).  Because of this property, the
contribution from triangle diagrams is absent in $\mathcal{M}_{+++}$; only
$\mathcal{M}_{++0}$ (and $\mathcal{M}_{--0}$, due to Eq.~(\ref{eq:sym1}))
contains such contributions.





\end{document}